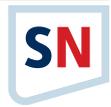

# Security Considerations for Internet of Things: A Survey

Anca Jurcut[1] · Tiberiu Niculcea[1] · Pasika Ranaweera[1] · Nhien-An Le-Khac[1]

## Abstract

Interconnecting "things" and devices that takes the form of wearables, sensors, actuators, mobiles, computers, meters, or even vehicles is a critical requirement for the current era. These inter-networked connections are serving the emerging applications home and building automation, smart cities and infrastructure, smart industries, and smart-everything. However, the security of these connected Internet of things (IoT) plays a centric role with no margin for error. After a review of the relevant, online literature on the topic and after looking at the market trends and developments, one can notice that there are still concerns with regard to security in IoT products and services. This paper is focusing on a survey on IoT security and aims to highlight the most significant problems related to safety and security in the IoT ecosystems. This survey identifies the general threat and attack vectors against IoT devices while highlighting the flaws and weak points that can lead to breaching the security. Furthermore, this paper presents solutions for remediation of the compromised security, as well as methods for risk mitigation, with prevention and improvement suggestions.

## Introduction

Internet of things (IoT) is the future of the Internet that will interconnect billions of intelligent communicating 'things' to cater diverse services to Information Technology (IT) users on a daily basis [92]. The IoT continues to affect the whole aspects of one's private and professional life. In the industrial sector, for example, smart devices will evolve to become active contributors to the business process improving the revenues of equipment manufacturers, Internet-based services providers, and application developers [3]. The IoT security is the area of endeavour concerned with safeguarding connected devices and networks in the Internet of things environment.

As IoT devices are interconnecting at every level and everywhere, interacting with each other and the human beings, it is evident that security takes the spotlight. Securing these devices will become everyone's priority, from manufacturers to silicon vendors (or IP developers), to software and application developers, and to the final consumer, the beneficiary of the security "recipe" that will accompany these IoT products. Together, they need to adapt to the market demands, innovate and improve processes, grasp new skills and learn new methods, raise the awareness, and elaborate new training and curricula programs.

The wearables are a hallmark of IoT, with designs that incorporate practical functions and features. From health to fashion and fitness-oriented devices, wearables make technology pervasive by interweaving it into daily life [105]. The main goal of these apparatus is to gather data such as heartbeat, burned calories, body or environment temperature, and so on and send it to the user for information purposes [8]. The wearables need to store the data locally or to the cloud, to generate historical reports about the achieved progress of the user.

Undoubtedly, the smart home collects as well an enormous amount of private information. For example, it may store the records about the absence or movements of the inhabitants, the temperature levels of the house in different

✉ Anca Jurcut
anca.jurcut@ucd.ie

Tiberiu Niculcea
tiberiu.niculcea@ucdconnect.ie

Pasika Ranaweera
pasika.ranaweera@ucdconnect.ie

Nhien-An Le-Khac
an.lekhac@ucd.ie

[1] School of Computer Science, University College Dublin, Dublin, Ireland

rooms, the water and electricity usage, and so on [139]. Much like the emerging smart homes, the smart office or smart building automatically controls energy-consuming devices such as heaters and lights to achieve a better efficiency without human intervention or micromanaging [131].

Smart cities use IoT devices like the connected sensors, lights, and meters to collect and analyse data for further usage in improving the infrastructure, public utilities and services, and much more [49]. The use cases are countless, but arguably the most important implementation is the smart grids, which helps tremendously with resource conservation [19]. In the smart healthcare domain, IoT technologies have many applications, and some of them are the tracking of objects and people, including patients, staff or ambulance, identification of individuals based on pervasive shared biometrics, and automatic data gathering and sensing [141].

The industrial Internet of things (IIoT), known as Industry 4.0, revolutionizes the manufacturing by enabling the addition and accessibility of far greater amounts of data, at higher speeds, and a lot more efficiently than before [16]. IIoT networks of smart devices allow industrial organizations to open big data containers and connect people, data, and processes from the factory floors to the offices of their executive leaders. Business managers can use IIoT data to get a full and accurate view of their enterprise health, which will assist them to make better decisions.

The IoT is also revolutionizing the supply chain management (SCM), a foundational business process that impacts nearly every enterprise [114]. Some of the possible use cases for SCM are asset tracking and fleet management. Asset tracking is possible based on radio frequency identification (RFID) tags or subscriber identity/identification module (SIM) cards with global coverage. This facility allows a supply chain manager to locate where a product, truck, or shipping container is, at one given time. Also, the fleet management enhances operators to know whether asset reliability, availability, and efficiency are all optimized.

The Internet of things is present at every level and sector of the society and will be even more rooted in, to become the new everyday normal. As IoT is everywhere, so should privacy and security be, inbuilt from the schematics of a product designer, until the last technician to influence in a way or the other, the finite apparatus. These devices undoubtedly will allow humans to become more efficient with their time, energy, and money in ways that are easy to forecast. Still, the lack of proper security frameworks and safeguards could lead to privacy being compromised and valuable data exfiltration to become possible. The convenience that IoT products and services bring to the lives of individuals has its price tag, and it could turn out to be a high bill in the end if security is not taken seriously by all the players of the IoT ecosystem.

### Contribution

This paper addresses some of the trending problems in the IoT, such as the ineffective identity, access, and trust management, by presenting solutions that are available in the market. The review of the most common threats and attacks raises the awareness about the importance of security, whereas exploring the reasons for safety breach boosts the understanding about why the IoT devices are still vulnerable. Depending on the fault tolerance capabilities of the apparatus in the aftermath of an attack, the remediation is not always possible, leading to the immediate replacement of the device for a new one. The operation is costly, labour-intensive, and time-consuming. Therefore, risk mitigation needs to be considered by everyone playing a role in the market. Mitigating risk starts with preventing the threat from happening. This survey offers guidance for threat and attack prevention by:

- showing how to raise the level and posture of security
- describing best practices for product design, manufacturing and development
- advising the consumers and lawmakers to be security-minded
- proposing a new design: Another important step in the reduction of the risk is to innovate and seek improvement.

This research proposes a new design, with mentions of disruptive technologies in order to replace the usage of the IT-related system and network models in the IoT ecosystem. The study elaborates as well on the issues posed by the scalability, complexity, and management of the IoT networks and identifies solutions for addressing it. With the advancement of quantum computing, big data and artificial intelligence (AI), predictive data analytics plays an important role not only in forecasting the future maintenance or the need for process optimization, but also in identifying device security weaknesses, data breach, and future possible attacks, before they even happen.

### Paper Organization

Rest of the paper is organized into five sections. Section " Related Work" summarizes the related surveys and researches that focus on IoT security aspects. In that section, we have attempted to classify the material under general, identity management, access control, and trust management. General IoT threats and vulnerabilities are presented in Sect. "General IoT Threats, Attacks, andVulnerabilities" where a summary of the threats and attacks are tabulated in Table 1. Our main contribution of this paper is mentioned in Sect. "Risk Mitigation" that includes the subsections on risk prevention and security improving practices. Section "Discussion" discusses the overall contribution of the paper, while Sect. "Conclusions" mentions the concluding remarks to the paper. The overall structure of the paper is depicted in Fig. 1.

## Related Work

While reviewing the existing work on the IoT security, a few research papers were chosen as relevant to this study and synthesized within this section. Looking at the market trends and developments, one can notice that there are still concerns with regard to security in IoT products and services.

Zhao et al. [154] conducted a survey on IoT security that expounds security issues related to the three-layer structure of IoT. The three layers of perception, network, and application are investigated against information, physical, and management security. As perception layer issues, node capture, fake nodes, malicious data, denial of service (DoS), timing, routing threats, side-channel attacks (SCAs), and replay attacks are identified. Similarly, network layer and application layer security issues are presented, while adoptable security measures are mentioned for each layer to mitigate the risks.

Ammar et al. [5] surveyed IoT frameworks on the emphasis of security and privacy. This paper clarifies the proposed architecture, and hardware sepcand points out the security features for 8 IoT frameworks. The considered frameworks include Amazon Web Service (AWS) IoT, ARM mbed IoT, Azure IoT suite, Brillo/Weave, Calvin, HomeKit, Kura, and SmartThings. Authentication, access control, communication, cryptography aspects of security are compared with these novel platforms. This is a comprehensive survey that provide valuable insights to IoT developers in selecting the most suited platform for their application.

Yang et al. [148] conducted a survey that covers the segments: limitations of IoT devices and their solutions, classification of IoT attacks, authentication and access control mechanisms, and security analysis of different layers. The paper identifies the battery life, and high-level computations required for employing strong cryptographic primitives are the main limitations of IoT devices. As solutions, energy harvesting and utilizing light-weight security protocols are proposed. Various existing IoT authentication schemes and architectures are presented, while security in perception, network, transport, and application layer are discussed.

Lin et al. [96] presented an overview of IoT system architecture, enabling technologies, security, and privacy issues, while discussing the integration of IoT with edge/fog computing platforms for various applications. Authors are distinguishing cyber-physical systems (CPSs) with IoT stating that CPS is a vertical architecture that forms separate CPS systems that do not interconnect, while IoT is presented as a networking infrastructure that interconnects various systems for resource sharing, analysis, and management. Confidentiality, integrity, availability, identification/authentication, privacy, and trust are discussed as security features of IoT. Moreover, possible security attacks for different layers are presented, while privacy aspects of IoT are presented for data collection, data aggregation, data mining, and data analytic cases.

Granjal et al. [53] conducted a comprehensive survey for analyzing existing communication protocols to identify security requirements in the intent of securing the communication channels. Protocols available for physical (PHY), media access control (MAC), network/routing, and application layers were extensively analysed for their security standards to derive security requirements. Among those, IPv6 over low-power wireless personal area networks (6LoW-PAN) and routing protocol for low-power and lossy networks (RPL) protocols were investigated thoroughly due to their wide adaptability in future IoT applications. Moreover, open research challenges are addressed in accordance with the identified security requirements.

Kliarsky [82] reviewed the existing threats, vulnerabilities, attacks, and intrusion detection methods that apply to IoT. The Open Web Application Security Project (OWASP) was identified as a trusted source to be informed of common threats and vulnerabilities. OWASP has published a list of what it is considered to be the top IoT vulnerabilities and mentions username enumeration and weak passwords as the top vulnerabilities. The paper referred the IoT Reference Model published by Cisco (presented in [29]) to identify possible attacks at every layer and then depicts the IoT communication stack by looking at some common IoT application and link layer protocols and technologies. Further, modus operandi and detection of intrusion for network assaults like the Mirai IoT botnet, denial of service (DoS), and routing attacks were presented. According to the paper, challenges that affect an IoT intrusion detection system (IDS) deployment include encryption, IPv6, scalability and management, and the complexity of the deployment.

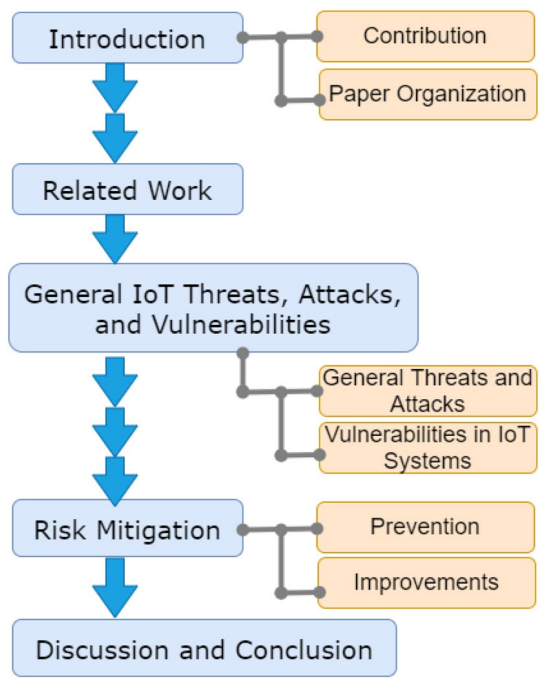

**Fig. 1** Structure of the paper

Rivas [128] explored the possibilities to secure the private IoT home network and presents means of network and IoT exploitation. The author mentioned that network design flaws, backdoors, DoS, spying, and man-in-the-middle (MitM) attacks are the other ways of compromising a network. The paper presented some of the core network services required to raise the security posture such as, Dynamic Host Control Protocol (DHCP), Domain Name System (DNS), Dynamic DNS, installation of intrusion detection and prevention systems (IDPS), proxies, and filtering. The paper pointed out that keeping an up to date inventory of the running systems of the connected devices in the network reduces the number of false positives on the IDS and filters out the protocols, ports, URI, sources, destinations, and applications. The inventory could be kept accurate by executing active or passive scans of the network from time to time.

Abomhara et al. in their paper [1] contributed to a better understanding of threats and their attributes by classifying the types of threats, analyzing, and characterizing the intruders and the attacks against IoT devices and services. Data confidentiality, privacy, and trust are three key problems with IoT devices and services identified by their research paper. The research concludes that it is important to consider security mechanisms for access control, authentication, identity management, and a trust management scheme, from the early product development stages.

Pawar et al. [118] uncovered the "Sybil attack in Internet of Things", by analysing the types of assaults according to Sybil's attacker capabilities, as well as some defensive schemes. The schemes include social graphs, behaviour classification, and mobile Sybil detection. The authors argued that the vulnerability of IoT systems in front of Sybil attacks leads to the systems generating wrong reports, spamming the users, spreading malware, and phishing websites, resulting in compromised privacy and private information loss. In addition, this paper proposed an enhanced algorithm to increase the detection of Sybil accounts by grouping similar user clickstream into behavioural clusters and by partitioning a similarity graph to capture the time distances between clickstreams sequences. Their study concluded that clickstream models are a powerful technique for user profiling and that future work needs to be done on the clickstream models to be able to detect: malicious crowdsourcing workers, forged online reviews about travelling related products, and identifying new methods of image-spamming attacks. The work of the authors is valuable to the present survey as it raises awareness about another type of attack on the rise, threatening the Internet of things products and services ecosystem.

Ouaddah et al. [112] conducted a survey on access control models, protocols, and frameworks in IoT. This survey analysed the security and privacy preserving objectives of scalability, usability, flexibility, interoperability, context awareness, distributed, height heterogeneity, light-weight, user driven, and granularity against the existing access control mechanisms. Role-based access control (RBAC), attribute-based access control (ABAC), Extensible Access Control Markup Language (XACML), capability-based access control (CapBAC), usage control (UCON), User-Managed Access (UMA), and OAuth methods are analysed to identifying the challenges in adopting access control schemes for IoT.

# General IoT Threats, Attacks, and Vulnerabilities

## General Threats and Attacks

An attack is an attempt to destroy, expose, alter, disable, steal, or gain unauthorized access to an asset [123]. An IoT attack is not peculiar from any typical perpetration conducted against an information system asset. The simplicity and scale of attacks are varied for IoT circumstances, where millions and billions of devices are potential victims for cyber-attacks on a larger scale.

An advanced persistent threat (APT) is a complex set of stealthy and continuous computer hacking processes, conducted by a person or a group of individuals targeting a specific entity [25]. An APT attack is aiming at stealing high-value information in business and government organizations, such as manufacturing, financial industries, and national defence [54].

Data and identity theft is another category of attack that gives grave consequences for the victim. As an example, the Google Nest thermostat was hacked via a USB connection within 15 s, in a show-off demonstration during the USA Black Hat conference in 2014 [67]. This attack scenario leads to privacy and consumer behaviour leaks, thus transforming the IoT device into a spyware.

The Mirai botnet attack was a botnet distributed denial of service (DDoS) attack perpetrated employing tens of millions of unprotected IoT devices to disrupt the operations of major Internet Service Providers (ISPs) [84]. This attack revealed the vulnerabilities of IoT devices proving their insecurities. The majority of the unknowingly recruited bots were millions of webcams. One of the after effects of this attack is that security needs restoration on these webcams and even replacing the cameras, as a final solution.

Ransomware is one of the top competitive online threats, leading to significant revenue loss for the companies infected [135]. It is becoming the most successful cyber-based attack because victims are willing to pay the demanded sum to regain the access to their private data. Even an adversary with malicious intent that do not possess a technical background to create a ransomware on their own could purchase a ransomware package from the dark web. WannaCry, CryptoLocker, CryptoWall, Petya, Locky, and TeslaCrypt are some of the frequently used types of ransomware [102]. IoT-based healthcare devices and services could become an attractive target for ransomware due to their handling of private medical stats.

SCAs are a type of attack that is arduous to mitigate with conventional means as they are exploiting the vulnerabilities of IoT devices that solely relies on the manufacturers ability to predicting flaws in their system [154]. Adversaries are focusing on time consumption, power consumption, or electromagnetic radiation emitted from the devices. Thus, shielding devices from such mishandling require more research, development budget, and time, factors that a typical IoT device manufacturer might not willing to invest in.

IoT devices are prone to man-in-the-middle (MitM) attacks [107]. A possible attack scenario would be in an instance where IoT device is communicating with the cloud for execution instructions, administrative decision making, or firmware updates. An adversary could attempt to redirect network traffic with an attack conducted at the network level, to include Address Resolution Protocol (ARP) cache poisoning or Domain Name System (DNS) modification attacks [62]. A self-signed certificate or tools such as SSLstrip can help attackers intercept Secure Hypertext Transfer Protocol (HTTPS) connections [28]. An example of MitM attack was the reported hacking of a Jeep Cherokee by a team of two ethical researchers [127]. Security vulnerability existed in the Uconnect dashboard computer of the car, causing a recall of 1.4 million vehicles. Table 1 summarizes the threats and vulnerabilities discussed in this subsection.

**Vulnerabilities in IoT Systems**

Unlike any traditional IT environment where systems are separated from the rest or each other by proper physical security, things in IoT are fixed and unattended. That makes the IoT systems more prone to tampering in terms of hacking. Companies need to ensure that data collection, storage, and processing would be continuously secure. It is required to adopt a new strategy in defence and encrypt data at each stage. Lack of local data encryption could lead to product hacking via physical tampering. Having physical access to a device allows an attacker to alter configuration settings in the cases of issuing a new device pairing request, resetting the device to factory settings, generating a new password, or installing custom fabricated Secure Sockets Layer (SSL) certificates to redirect traffic to another server owned by them.

In cryptography, the terminology of a weak key refers to the key phrase that is used with a specific cryptographic algorithm or a cipher that is exposed with brute force (exhaustive key search), or guessing. Weak keys usually represent a tiny fraction of the overall keyspace, the set of the whole possible permutations of a key. They are very unlikely to give rise to a security problem. Nevertheless, a cipher should employ a key with a appropriate length. The key size or the key length is the number of bits found in a key and used by a cipher. In practice, cumbersome long keys are utilized for modern cryptography for achieving computational security, so that breaking the cryptosystem is computationally infeasible. Though, the advent of quantum computing proves otherwise. The algorithms that are used for cryptosystems are either symmetric [e.g. Advanced Encryption Standard (AES)], asymmetric (e.g. RSA), or hybrid (combination of both symmetric and asymmetric) [78]. Such cryptoalgorithms are linked to the weakness of a key. Depending on the used algorithm, it is common to have various key sizes for the same level of security. As an example being the security available with a 1024-bit key using asymmetric RSA considered to be approximately equal in security to a 80-bit key from a symmetric algorithm [134].

Table 1 Summary of general threats and attacks possible for IoT

| Threat/attack | Description | Consequences | References |
|---|---|---|---|
| Advanced persistent threat (APT) | An adversary targeting an information system that launches continual hacking attempts | Complete control of the hacked system and its assets | [25, 54, 75] |
| Data Identity Theft | A hacking attempt launched with actual user credentials as an impersonation attack | Privacy leakages | [67] |
| Distributed denial of service (DDoS) attack | DoS attack launched from multiple locations simultaneously | Service interruption due to overloading | [75, 84] |
| Botnet attack | A network of bots acting to compromise a singular or multiple targets | A DDoS attack | [84] |
| Ransomware | A malware once installed to a system demands a ransom (typically financial) from the owner | Denied access to a part or entire system or threatens to publish sensitive information until the ransom is settled | [102, 135] |
| Man-in-the-middle (MitM) | Attempting to access the traversing information of a communication link in between the sender and the receiver | Revealing the information and protocol, injecting false/malicious content | [31, 73] |
| Side-channel attack (SCA) | Analyses a physical property of a device via tampering | Information, keys, or even a protocol could be revealed | [75, 154] |

One popular and comfortable method for users to interact with an IoT device is via a web browser or a smartphone app. Sometimes, devices with a more processing power run a small web server that allows the user to use a web-based graphical user interface (GUI) to send commands. Other devices offer the user the possibility to interact with them via their application programming interface (API). When the user wants to send commands to a device or control it remotely, they open an inbound port on the router via a Universal Plug and Play (UPnP) request. The lack of encryption is one of the major privacy concerns. Devices can pass private data, login credentials, or tokens in plain text, letting an attacker intercept them via a network eavesdropping technique. Cryptographic protocols are required to ensure the security of both the infrastructure itself and the information that runs through it [72]. Moreover, the design of such protocols should be robust enough to resist attacks [70, 71, 74] and must be tested for their functional correctness (i.e. application of formal method) before they are used in practice [69, 86].

One of the communication protocols prone to eavesdropping is Telnet [142]. The protocol was developed long before the Internet took shape, in a time when not much consideration was given to data confidentiality while in transit. The whole data transmitted with this protocol is susceptible of being intercepted. Hypertext Transfer Protocol (HTTP) is another example of insecure communication protocol still in use, which empowers an eavesdropper to view the communication between a client and the server [20]. Although the attacker is not able to capture the password from the web server, they are capable of harvesting other types of data, such as accurate information about the configuration or even a valid cookie that will allow them to impersonate a legitimate user and then gain access to the administrative interface of the firewall. Simple Network Management Protocol (SNMP), v1, and v2c are insecure protocols which expose a firewall for complete reconfiguration in the read-only mode. The File Transfer Protocol (FTP) and Trivial File Transfer Protocol (TFTP) are used to copy files from/to a device to update the system configuration or software/firmware. Compared to TFTP, FTP provides the mechanisms for authentication [104]. Still, both protocols transmit the data in an unencrypted manner and are therefore susceptible to an eavesdropping attack.

The scope of developing products following the minimum viable product (MVP) technique is to build a product fast and release it on the market to learn about customer reactions [109]. A new version of the product lands on the designing workbench, soon after gathering the feedback from the previous release ends. The tremendous pressure to release the MVP in a short amount of time leads to neglecting the security and privacy of the final product. Moreover, "ship and forget" mentality of some manufacturers leaves the customers with devices that are running several years' old software that were never updated. Thus, such devices have severe security flaws. On the contrary even if an update is available, the vast majority of the typical customers do not have the skills, energy, willingness, or time to go through the hassle of updating their IoT devices. No matter what manufacturers do, sometimes the customer still is the weakest link when it comes to securing various IoT devices.

It is a challenge for IoT companies to agree on interoperability protocols and standards for the sharing and protecting of data. Competing standards, proprietary devices, vendor lock-in, and private networks make it hard for devices to share a common security protocol. Embracing one IoT common standard by the companies is one of the barriers that hold back mass adoption of IoT security protection. Nonetheless, there are IoT standardization efforts. Samsung, Intel, and Cisco support the Open Interconnect Consortium (OIC) [46, 117]. LG, Microsoft, and Qualcomm back The Linux Foundation's AllSeen Alliance [94]. Google sponsors Zigbee and Thread Group Alliance, a UK-based Hypercat standard [116]. There are even more unifying efforts in the works that are industry specific to agree on a common networking protocol. Companies still have to conclude the battle for software standards. Gartner argues that the sheer sum of IoT use cases contributes to a wildly contrasting total of approaches to solving IoT problems, which creates interoperability challenges and, ultimately security gaps [14].

Devices connected across multiple geographies lead to practical issues of international enforcement when dealing with IoT. Country-specific privacy laws are insufficient as the reach of IoT data is global. Unless there are globally accepted laws which govern the usage of IoT information, data larceny will continue.

## Risk Mitigation

Mitigating the risk of an intrusion attempt or attack against an IoT device is not an easy thing to do. Having a higher degree of security protection at every level will discourage the attacker to pursue his goal further and make him give up in the end, by cause of the amount of effort and time needed versus benefits. Mitigation needs to start with prevention, by involving every actor in the market, from manufacturers to consumers and lawmakers, and make them understand the impact of the IoT security threats in a connected world. Another way to mitigate risk is to keep abreast of the times by improving and innovating, from the ground up, and by finding new methods and designs to outgrow the shortcomings of the market.

## Prevention

This subsection discusses the solutions that can be employed for prevention of the security threats in IoT systems, as illustrated in Table 2.

The honeypot system is the new weapon that required to be included in the cyber-security arsenal of the organizations to defend against attackers that try to penetrate secure networks through IoT back-doors [7, 35, 113]. A standard cyber-security defence should include the conventional prevention techniques along with the visibility to detect inside-the-network threats in real time, through identification of distinctive threats and their levels by setting up an incident response playbook to remediate infected systems. The ThreatMatrix platform provides a form of risk detection for various categories of dangerous vectors including ransomware, phishing, stolen credentials, and reconnaissance attacks [65]. The matrix is customized to fit into various landscapes, which creates a trap out of each IoT network. The Attivo Networks IoT solution protects widely used protocols such as Extensible Messaging and Presence Protocol (XMPP), Constrained Application Protocol (CoAP), Message Queuing Telemetry Transport (MQTT), and Digital Imaging and Communications in Medicine (DICOM) servers which are used by the IoT vendors to support a wide set of applications that allow for more excellent machine-to-machine communication and monitoring, concerning critical data and machine status [11]. The Attivo analysis engine is capable of analyzing the techniques used in the attack, the lateral movement of the assault, what systems are infected, and will provide the necessary signatures to stop the attack. Analyzing the attack improves incident response skills and capabilities, by automatically or manually blocking and quarantining the attack through integration with third party systems and solutions for intrusion and prevention.

As IoT market will mature, the general public can access new professional training, and University taught programs. Awareness and proper training is paramount for owners of the smart devices to understand how to implement some basic security countermeasures that are the first and the best line of defence [10, 63, 140].

Manufacturers know the best application and intention of their products. They do not get the direct feedback from the owner. Many devices include open-source software as part of the code that is running on the product. Device manufacturers need to maintain lists of open-source components that are used in the production process. When the community identifies a vulnerability in one of those components, an update can be made available quickly to the device owners. Also, device manufacturers need to ensure that communication procedures are established with the device holder to allow immediate responses in case these vulnerabilities arose [108].

Chipsets are the core of the device, and IoT devices make no exception [22]. The better designed is the chip, the more secure it is and harder to crack when compared to a software solution that promises to offer the same functionality. Over the past five years, silicon suppliers have had to complement their offering with a full-fledged featured software stack to support their silicon, and hence, moving beyond hardware drivers, into network and security stacks or even embedded operating systems. Atmel Microchip, for example, is putting the accent on hardware security, by developing world-class embedded security solutions to ensure trust for each system design [106].

At the application level, organizations that develop software need to be writing code that is more stable, buoyant and reliable, with better code development standards, training, threat analysis and testing. Application developers will have to team-up with application penetration testers to analyse the logic and operation of exposed applications, as an attacker would do in his attempt to gain access to sensitive data or to bypass logic controls and compromise a system. It is of high importance to repeatedly test for resistance against attacks since new ways of assault are developed even after a product or solution is created and released. In addition to testing in development and quality assurance phases, testing IoT systems in production settings is highly recommended. Extreme physical operating conditions do not have to be the only test that devices are subject to, but also to extreme computational conditions, which include resistance to denial of service (DoS) and jamming-style attacks where a flood of information hits the product to attempt and confuse, overpower, or disable it [32].

Static analysis and source code reviewing practices do not detect risks and vulnerabilities from penetration testing alone. Organizations and developers need to define flexible security architecture and deploy data-centric security technologies to support speed, agility, cost-effectiveness, and innovation, in a highly connected world. For traditional IT ecosystems, various systems development life cycle (SDLC) methodologies have already been put in place and proven to be successful in guiding the processes involved to create a software component that easily integrates with other software components [143]. Developing for IoT is not very different and should address all the stages, from design and development to testing and debugging, to deployment, to management, and to decommissioning. For developers of IoT with the mobile client, Cloud or IoT applications finding the right strategy and solutions are not an easy task. The mission of the developers is not only to bring these solutions to market rapidly but also to ensure that appropriate security and data protection measures are implemented from the beginning because no business can afford the high costs in the aftermath of data theft. Improper security system exposes confidential and valuable customer information, financial transactions, and mission-critical operational data, and hence, lowering the risk of data exfiltration needs to be at the core of their activities [15].

Table 2  Summary of security risk prevention methods for IoT

| Risk prevention method | Description | Benefits | Tools/sources |
| --- | --- | --- | --- |
| Honeypots | A mechanism that lays a trap on adversaries who are intending to perform unauthorized acts | Detection and counteracting threats without affecting the information system | ThreatMatrix, Attivo Networks |
| Raising awareness through training | Introducing professional training programs for IoT users and developers | Users will follow secure practices while awareness is raised for overcoming general security attacks such as Phishing attacks | Perpetual Solutions, MCU Solutions, cybrary |
| Immediate response to detected vulnerabilities | Quick generation of updates and patches to detected flows, specially in open-source software | preventing exploitation of well advertised vulnerabilities in case of open-source material | [36] |
| Security on Chip | Integrating security for IoT chips/hardware at the manufacturing stage | An extra layer of security, that adds faster response due to the implementation at the perception domain | Atmel micro-chip, [85, 115, 126] |
| Exhaustive security testing | Better testing schemes should be introduced to cover penetrative, access, physical, computational aspects | Maximum assurance granted before releasing the IoT device to the market | [23, 56, 98] |
| Security on SDLC | Security practices should be integrated into the software design stages in SDLC | Security is addressed as a main goal of all the software products with improved compatibility and inter-operability | [40, 51, 122] |
| GDPR | Legislation's put forward by the EU for data protection | Improved awareness of the general public and the availability of a legal framework for accountability of digital and privacy violations | [39] |

Consumer's education starts with best practices provided by the organizations selling the product [137]. The highly efficient ones include regularly changing passwords, which is still among the frequent causes for a security breach and also offering advice on the safety patches and updates. Consumers need a level of confidence and comfort if they are going to buy IoT products. They trust the manufacturer's brand to guarantee some degree of design and quality. When a consumer values security, they will insist that the goods they buy are secure and will pay the price that comes with it.

The European Union released new guidelines on how companies operating in Europe have to handle and protect the data of their customers. As of 25 May 2018, organizations need to comply with this General Data Protection Regulation (GDPR) [145]. The GDPR introduces developments in some areas of EU data protection law. They will have a direct impact on the way product manufacturers, application developers, social platforms, and other entities involved in the IoT field and especially design and bring to market IoT-based devices, systems, and applications.

The GDPR imposes obligations on data controllers to adopt significant new technical and organizational measures to demonstrate their compliance [39]. These include conducting data protection impact assessments in certain circumstances which are likely to arise in connection with IoT systems. The GDPR will confer new substantive rights of data subjects about their private information. These substantive rights include an express right to be forgotten, the right to object to automated decision making, and data portability rights. The design and engineering of IoT devices, applications, and systems will need to accommodate the necessary capabilities to facilitate the exercise of these rights in compliance with the GDPR, particularly about data portability.

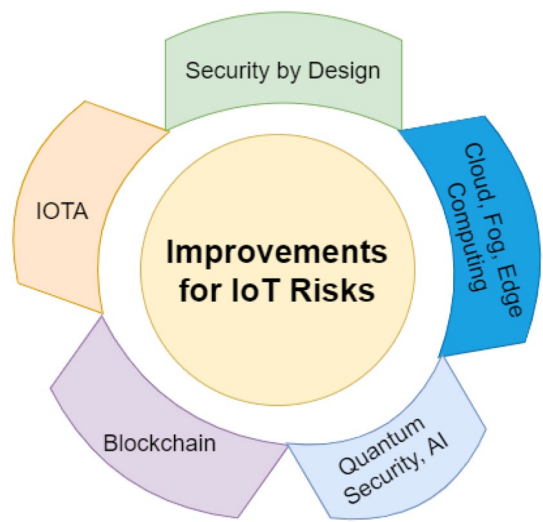

Fig. 2 Possible improvements for IoT risk mitigation

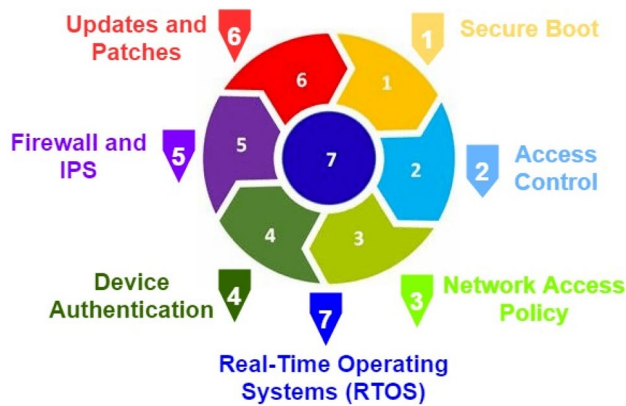

Fig. 3 Security by design approaches

## Improvements

This subsection addresses solutions that can be employed for improving the security of IoT systems, as illustrated in Fig. 2.

### Security by Design

In embedded systems such as gateways, hubs, and similar network entry points for devices and things that connect to them, there is a need for a different approach to be considered when improving security, which starts in the early planning of the product with security by design (SbD) concept as depicted in Fig. 3.

- Secure Boot

Security practitioners need to build a multilayered approach to IoT ecosystem right from initial secure booting to establishing trust and integrity of the software on the IoT device. To establish these, the role-based access control (RBAC) makes sure that users access only those privileges and applications that they require as part of their job role [111]. Also, incorporating principle of least privilege, persistent device authentication and building proper host-based firewalls and deep packet inspection capability will enhance the trust and integrity [41, 77]. This deep integration of interconnected devices that embed into our daily lives means that security is of paramount importance. Applying add-on security controls to each IoT device is impractical and wasting resources. Security needs to be inbuilt, fitting the environment and supporting system functionality without restrictions.

When the System-on-Chip (SoC)-based devices boot its system, authenticity, and integrity of the software, firmware and hardware components are checked with different means. The ways to ensuring secure booting and verifying integrity of the installed software and firmware are important for guaranteeing its reliability in the context of marketing [66]. Methods such as Elliptic Curve Digital Signature Algorithm (ECDSA), Secure Hash Algorithm (SHA), direct memory access (DMA), and physical unclonable function (PUF) are employed for secure booting and remote attestation [58, 68]. Embedding these methods for boot loading processes is mitigating attack scenarios plausible with malicious boot agents. As such, the groundwork of trust settles, but the device still needs protection from various run-time threats and malicious intentions.

- Access Control

The operating system's built-in access controls, mandatory or role-based, have the benefit of managing the privileges for the device components and applications so that they only access those resources assigned to them. In the case of an intrusion, access control ensures that the intruder has minimal access to other parts of the system. Device-based access control mechanisms are similar to the network-based access control systems such as Microsoft Active Directory [5]. If someone manages to steal corporate credentials and gains means of entry to the network, the access to such compromised information restricts to only those segments of the network, authorized by those appropriate credentials. The principle of least privilege commands that minimal access required to perform a function need to be permitted, to minimize the effectiveness of a breach of security [77].

- Network Access Policy

Once the enterprise network incorporates IoT devices, the IT organization has to create or alter the configuration of the network access policy as part of a corporate policy enforcement strategy. This strategy needs to determine whether and how these devices connect to the network, maybe separated into virtual segments, as well as what role they will be assigned that will govern their access. Some of the advantages of network segmentation are improved security, performance boost, and network problems isolation [24]. By creating network segments for IoT devices only, the principle of least privilege is applied, thus limiting further movement across the network for cyber-criminals with unauthorized access. Network performance improves by isolating IoT transactions to a defined segment, which implies minimizing local traffic and in the end reducing network congestion. For a better isolation of a problem, access to the network can be handled by implementing another technique, called segregation [9]. Segregation works by combining virtual local area network (VLAN) and firewalls, where a set of rules is present and enforces to control which devices are permitted to communicate on that network segment in ingress and egress directions [89].

- Device Authentication

Device authentication needs to be triggered when the asset is added to the network for the first time, even before receiving or transmitting data. Embedded devices do not wait for users to input the credentials required for accessing the network, but their identification needs to happen correctly before authorization. Similar to how user authentication mechanism allows a user to access the corporate network with a username and a password, machine authentication allows devices to access the network with a pair of credentials stored in a secure storage area. These authentication mechanisms are mostly referred as device-to-device (D2D) authentication, where authentication credentials are exchanged through a machine-to-machine (M2M) channel [21, 59]. Resource constrained nature of IoT devices is encouraging lightweight approaches to maintain the transmission efficiency in a satisfactory level [2, 52]. Moreover, it will improve the operating time of the battery operated devices [138]. Thus, embedding a proper authentication protocol through circumspect designing is vital on both security and transmission perspective.

PUF is a nascent concept employed for D2D, M2M, IoT device, and even vehicular entity authentication. The idea of the PUF is to generate a unique identifier from a challenge response pair (CRP) that is derived from the unique features inherited by the circuitry over the fabrication process. The complexity and the secureness of the PUF based schemes are reliant on the number of CRPs associated with them [88]. In addition to authentication, PUFs can be employed for secure storing. New directives on PUFs can be found in [48, 59, 88, 100, 153].

- Firewalls and Intrusion Prevention Systems

The IoT devices require firewall and deep packet inspection (DPI) capability to control the traffic that is meant to terminate at the instrument [90]. Deeply embedded devices have various protocols, distinct from enterprise IT protocols, and a host-based firewall or intrusion prevention system (IPS) is highly required [37, 57]. As an example, the smart energy grid network has its proprietary set of protocols defining how devices talk to each other [61]. Protocol filtering and DPI capabilities, applicable to each industry, are required to identify malicious payloads hiding in non-IT protocols. The device should not bare itself with filtering higher-level, general Internet traffic, as the network appliances take care of that. But it does need to filter the specific data destined to terminate at the apparatus, in such a way that makes optimal use of the limited computational resources available.

- Updates and Patches

Once the device is operational, it starts to receive patches and software updates [95]. Devices need to authenticate the patches rolled out by the administrators, in a way that does not consume bandwidth or impair the functionality or safety of the apparatus itself. Contrary to how companies like Microsoft send updates to Windows users and tie up their computers, IoT products need receiving software updates and security patches in a way that conserves their limited bandwidth and connectivity and eliminates the possibility of compromising functional safety [43]. These devices are in the field, performing critical functions, and are dependent on the total of security patches that are available to protect them against the inevitable vulnerabilities of the wild. In the future, considering the increased numbers of devices and the expected frequency of updates, this work will transition from active participation by humans to automatic over-the-air update processing. Exception processing will become an isolated human intervention rather than handling and processing each update as it arrives, which suggests an increased level of monitoring and reporting on the status and progress of update processing across the inventory of gateways, routers, and devices involved [42].

- Real-Time Operating Systems

The majority of IoT appliances have common operating systems (OSs) that are incapable of addressing specific security requirements. These systems tend to be over-featured and geared with functionality that is useless for the connected devices. Also, there is not much importance given to fixing the various vulnerabilities caused by the poor design, bad implementation, or improper use of operating systems in these products. Building security in at the OS level takes the stain off device designers and developers and gives them more time at hand to configure systems to mitigate threats and ensure their platforms are safe. A real-time operating system (RTOS) is an operating system that manages the hardware resources, hosts applications, and processes data in real time [103]. RTOS defines the real-time task processing time, interrupt latency, and reliability for both hardware and applications, and in particular, for low powered and memory constrained devices and networks [83]. The main difference between RTOS and a general purpose OS stands in its high degree of reliability and consistency when measuring application's task acceptance and completion timing. RTOS is a critical component to building comprehensive embedded systems for IoT solutions for both consumer and industrial IoT [50]. More and more RTOS offerings are surfacing the IoT market and solutions like KasperskyOS, promise to bring a multitude of features to strengthen the security of the device [91]. Some of the main features guaranteed by RTOS are proprietary microkernel and a free security engine, multi-level compatibility, security domain separation, mandatory identification and labelling, and various policies enforcement [6].

**Blockchain**

IoT concept is in its development stages, but it is already offering technologies that allow for data collection, remote monitoring, and control of the devices. As it evolves, IoT transitions toward becoming a network of real autonomous devices that interact with each other and with their environment around them to make smart decisions without human intervention [87]. As such, the blockchain forms the foundation that will support a shared economy that works on M2M communications [155]. Blockchain technology leads to the creation of secure mesh networks, where IoT devices will interconnect while avoiding threats such as impersonation or device spoofing. As more legitimate nodes register on the blockchain network, devices will identify and authenticate each other without a need for central brokers and certification authorities. The network will scale to support more and more devices without the need for additional resources [132].

There are possible applications of blockchain technology in the context of IoT security. Blockchain hashes the device firmware on a continual basis, and if the firmware state changes by even a single digit by the cause of malware altering the firmware code, then the hash failure will alert the device owners [93]. To be able to send data or to check for new instructions, a device hashes the information it wants to send and places the hash into a blockchain. Then, the recipient of the package hashes the same data, and if the resulting hash matches the hash on the blockchain, then it means that the payload has not changed in transit. As each device has a blockchain public key, devices need to encrypt messages to each other employing a challenge/responses mechanism to ensure the device is in control of its identity; hence, it might be a useful idea to require a universal identity protocol for every instrument. Devices develop their reputation in the same way as Keybase key directory does, where each device has a public key [79]. Cryptographic reputation systems cover above devices. A certification agency for things which audits the device and generates an identity for it on the blockchain could be a solution. So once the instrument is historically born on the blockchain, the device's identity will be irreversible. For sensors such as global positioning system (GPS), temperature, and humidity, environmental inputs are unique to each other. This uniqueness in conjunction with the International Mobile Equipment Identity (IMEI) and Original Equipment Manufacturer (OEM) firmware hashes are forming a solution that is considered to be the ultimate in tamper-resistant device identification.

Furthermore, the blockchain technology can be used to promote digital business process without the need for a complex infrastructure [144]. These blockchain enabled interoperable platforms support companies to exchange authentication information with each other. The lack of shared identity stacks prevents companies from identifying and authenticating users with other businesses. With the blockchain technology, companies can keep stacks of common identities for user authentication through biometric data. Blockchain can support as well an interoperable ledger for identity exchange among multiple entities. From the cryptography point of view, the blockchain technology will set up the protocols for connectivity among devices through a biometric data validation process. The network running nodes will receive biometric data associated with respective devices and their time stamps. The network needs to confirm whether a device and a particular identity intersected each other within a time interval, to be able to authenticate a user.

As with each disruptive concept that turns into an effective offering, the blockchain model is not perfect and has its flaws and shortcomings [152]. Novel attack vectors such as forking attacks are creating a hassle for IoT service providers as blockchain was the security solution for achieving a privacy preserving service platform [146]. Scalability is one of the main issues, considering the tendency towards centralization with a growing Blockchain [30]. As the blockchain grows, the nodes in the network require more storage, bandwidth, and computational power to be able to process a block, which leads to only a handful of the nodes being able to process a block. Computing power and processing time is another challenge, as the IoT ecosystem is very diverse and not every device will be able to compute the same encryption algorithms at the desired speed. Storage of a continuously increasing ledger database on a broad range of smart devices with small storage capabilities, such as sensors, is yet another hurdle. The lack of skilled people to understand and develop the IoT-blockchain technologies together is also a challenge. The lack of laws and a compliance code to follow by the manufacturers and service providers is not helping both the IoT and blockchain to take off as expected.

### IOTA: The Post Blockchain Token

The launch and success of the Bitcoin cryptocurrency during the last years proved the value of the blockchain technology. However, as shown above, this technology has some drawbacks, which prevent its mass adoption as the only global platform for cryptocurrencies. Among these disadvantages, a particularly notable one is the limitations of making micropayments, which have increased importance for the rapidly developing IoT industry. Specifically, in the cryptocurrency systems, a user needs to pay a fee each time he initiates a transaction; hence, for a small amount, the fee might be many times larger compared to the transaction, and the transaction would make sense in the first place. These charges serve as an incentive for the creators of the blocks, and it is not easy to get rid of them. The existing cryptocurrencies are independent systems with a distinct separation of roles, for example, transaction issuers and transaction approvers. Such systems create inescapable discrimination of some of their elements which in turn creates conflicts and makes the entire collection of items to spend resources on conflict resolution. These arguments justify the search for solutions essentially peculiar from the blockchain technology, on which the Bitcoin and many other cryptocurrencies base their code.

IOTA is a disruptive transactional settlement and data transfer layer for the IoT [47]. At the foundation of IOTA, there is a newly distributed ledger, called the Tangle, which overcomes the inefficiencies of the blockchain design and introduces a new way, called directed acyclic graph (DAG), to reach consensus in a decentralized peer-to-peer system [44]. The users of IOTA automatically act as validators, allowing transaction validation to become an intrinsic property of utilizing the network. Each transaction requires that the sender verifies two previous transactions, which results in an infinite scalability, as opposed to the blockchain consensus design [45]. It enables people to transfer money without fees, meaning that even infinitesimally small nanopayments are possible through IOTA. The system could turn into the missing puzzle piece for the Machine Economy to emerge and reach its full desired potential. IOTA is meant to be the public, permission-less backbone for IoT that enable true interoperability between the devices.

## Cloud, Fog, and Edge Computing

Cloud computing and IoT build a couple that could work in a symbiosis. The growth of IoT and the rapid development of associated technologies create a popular connection of things that leads to the production of large amounts of data, which needs to be stored, processed, and accessed. This newly formed opportunity of cloud computing and IoT will enable new monitoring services and high processing of sensory data streams [17]. For example, cloud computing stores the sensory data, so that it is used later for smart monitoring and actuation with other smart IoT products. Ultimately, the goal is to transform the data into insight and drive productivity and cost-effective actions from this. The cloud plays the role to serve as the brain to improved decision-making and optimized Internet-based interactions. Cloud computing offers a realistic utility-based model that will enable businesses and users to access applications on demand anytime and from anywhere [4]. Amazon, Microsoft, and IBM are some of the major companies that are providing cloud computing services which have also incorporated offerings for the IoT market, like AWS IoT [18], Azure IoT Suite [81], and Watson IoT [55].

Infrastructure-as-a-Service (IaaS) provides the necessary hardware and software upon which a customer can build a customized computing environment. Computing and data storage resources, as well as the communications channel, are bound together with these IT resources to assure the stability of applications used in the cloud [119]. Symphony Link, offered by Link Labs, is a wireless solution for enterprise and industrial which connects IoT devices to the cloud securely [97]. The Symphony Link design is for Low Power Wide Area Network (LPWAN) applications, which are easily scalable and have high reliability.

In a Platform-as-a-Service (PaaS), a proprietary language is supported and provided by the platform's owner [119]. The platform eases communication, monitoring, billing, and other aspects, to ensure the scalability and flexibility of an application. Nonetheless, there are some limitations, regarding the programming model and supported languages, the ability to access resources and the long-term persistence. Other platforms like Wind River® Helix$^{TM}$ and ARMmbed IoT Device Platform provide a portfolio of software, technologies, tools, developer ecosystem, and cloud services for dealing with the challenges and opportunities at the system level, created by the IoT [80, 110]. These tools make the creation and deployment of commercial, standards-based IoT solutions possible at scale. Blockchain-as-a-Service platforms are starting to become popular due to its wide adaptability in Bitcoin and cryptocurrency applications, which is considered as a solid innovation during the last eleven years of its presence in the financial trading markets [132]. The application of this emerging technology is showing great promise in the enterprise.

Software-as-a-Service (SaaS) enables cost-effective value added services for many IoT applications that provision real-time data visualization and analytical support for its consumers [125]. These services mimic the application service provider (ASP) on the application layer. Usually, a specific company that uses the service would run, maintain, and facilitate support so that it assures reliability over an extended period. Device Authority's KeyScaler IoT IAM platform can assist in solving mass device provisioning, secure onboarding, certificate revocation and rotation, and solving credential management problems for Amazon Web Service (AWS)-based IoT customers [34]. This is an important step to take in securing IoT devices and their data. AWS IoT cloud platform lets connected devices to interact with cloud applications and other assets easily and supports a vast amount of messages to be processed and routed to AWS endpoints.

Although powerful, the cloud model is not the best choice for environments where Internet connectivity is limited or operations are time-critical. In scenarios such as patient care, milliseconds have fatal consequences [151]. As well in the vehicle-to-vehicle (V2V) communications, the prevention of collisions and accidents relies on the low latency of the responses [133]. Due to these novel requirements, cloud computing is not consistently viable for many IoT applications. Thus, it is replaced by the edge computing paradigms such as fog computing, mobile cloud computing (MCC), and multi-access edge computing (MEC) [38, 120, 124, 150].

Fog computing, also known as fogging, is a decentralized computing infrastructure in which the data, compute, storage, and applications split in an efficient way between the data source and the cloud [99]. Fog computing extends the cloud computing and services alike, to the edge of the network, by bringing the advantages and the power of the cloud to where the data arise initially as illustrated in Fig. 4. The main goal of fogging is to improve efficiency and also to reduce the quantity of data that moves to the cloud for processing, analysis, and storage. In fogging, data processing takes place in a router, gateway or a data hub on a smart device, which sends it further to sources for processing and storing that reduce the bandwidth payload towards the cloud. The back-and-forth communication between IoT devices and the cloud can negatively affect the overall performance and security of the IoT asset. The distributed approach of fogging addresses the problem of the high amount of data coming from smart sensors and IoT devices, which would be costly and time-consuming to send to the cloud each time. Fog networking complements the cloud computing and allows short-term analytics at the edge while the cloud performs resource-intensive, longer-term analytics [136]. Trends demonstrated that inexpensive, low-power processing, and storage are becoming more available and will drive the growth and usage of fog computing in IoT. Processing of data migrates even closer to the edge and becomes deeply rooted in the very same devices that created the data initially, thus generating even greater possibilities for M2M intelligence and interactions.

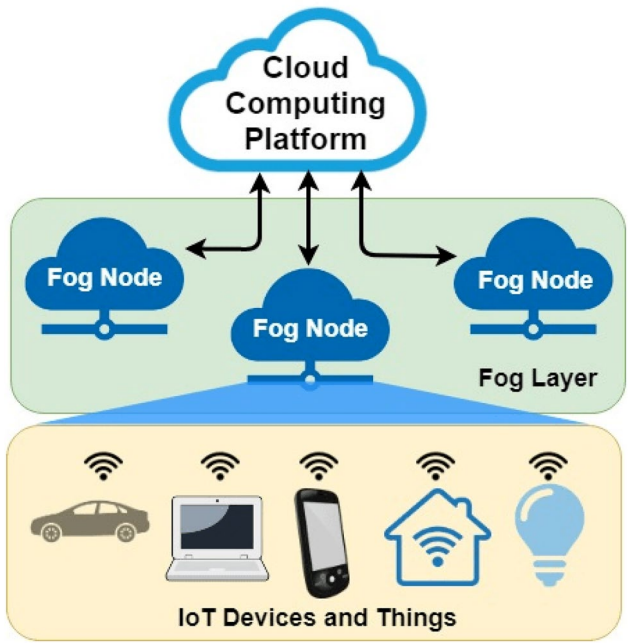

**Fig. 4** Extension of cloud services to the edge by fog computing

**Quantum Security, AI, and Predictive Data Analytics**

In quantum computing (QC), computations are handled faster than the classical computers which surpasses its capabilities with a considerable margin [121]. The QC allows for more data crunch with quantum speed and the ability to run an entire set of inputs at the same time, thus getting instant results. Security experts are predicting that quantum cryptography will replace the existing security solutions in all digital systems that are prone to data hacking, including national defence, finance, self-driving vehicle industry, and the IoT, with the potential to be unhackable [130]. Quantum computers will become a technological reality sooner than expected, and it is vital to study the cryptographic schemes used by adversaries with access to a quantum computer. Post-quantum cryptography is the study of such plans that arose from the fundamentals of popular encryption and signature schemes [26]. Existing elliptic curves and Rivest–Shamir–Adleman (RSA) algorithms can be broken using Shor's algorithm on a quantum computer via factoring and computing discrete logarithms [13, 129]. Though, schemes such as McEliece, lattice, hash, code, multivariate, and super-singular elliptic curve isogeny methods are envisaged to develop Quantum Resistant (QR) security systems [12, 27, 64, 101].

Quantum encryption methods are being engineered by embedding quantum mechanics on microchips/processors to enhance the security of random number generation in cryptographic protocols [33]. The security of cryptographic protocols is dependent on the randomness of the keys. At present, the vast majority of these protocols use algorithmic pseudo-random number generators. The approach followed by [149] could be employed for revolutionizing randomness in existing security and communication protocols to prevent hacking and guessing attempts.

In the IoT ecosystem, the volume of data and also the data types are increasing. Data comes from a wide variety of sources. It is obvious that the conventional computing systems cannot handle the amount of data generated from IoT based sensors and meters serving myriads of services and applications. The method of predictive analytics is facilitating the decision makers to sort and understand the type, amount, and frequency of data to be expected, so that they can take immediate actions [76]. The precision of the prediction method is reliant on the amount, variation, and duration of data. Predictive data analytics will be a core solutions to provide close-to-zero downtime for many sectors; especially, industrial automation. Prevention of failures occurring on mission-critical devices and forecasting the domino-effect originated from the incident is plausible with predictive analytics performed on IoT systems. Security-wise, it is capable of discovering a data breach before it happens. Predictive data analytics will be supported by machine learning approaches executed on edge, without the requirement of connecting to the Internet. In a smart city, various systems such as traffic system, lights, motion sensors, closed-circuit television (CCTV), meters, utilities, and smart buildings exist. QC can potentially handle the verification and the validation process faster across every system and ensure continuous optimization for these systems.

Given the new data and scenarios, artificial intelligence (AI) and IoT are shaping up to be a symbiotic pairing, where AI depends and thrives upon high data inputs that IoT delivers [60]. Cognitive systems of AI evolve and improve over time, inferring new knowledge without being explicitly programmed to do so. Another way that AI can pair up with IoT technologies is by bringing cognitive power to the edges of IoT, through embodied cognition [147]. That means AI capabilities are placed in an object, avatar, or space (such as the walls of a spacecraft), enabling it to understand its environment, and then reason, and learn. These objects may have the ability to interact in more natural human-like ways, such as written and verbal communications and gestures, with the observations of actual humans living and working in their proximity.

## Discussion

Our survey unveils concerns over some outstanding issues of IoT ecosystem. The most relevant are the management of the identity, access control, and trust towards IoT products and services. Ineffective authentication methods introduce a trust deficit across IoT network gateways, which expose these devices and their data to perpetrators. Another point in question is the use of centralized, traditional IT computing systems, and network models in an IoT environment that are meant to be self-governed and decentralized. IoT belongs to the new era, and every actor that has a role to play in this environment needs to adapt to the requirements of this new ecosystem. These systems contain continuously growing, huge number of devices, and the scalability, complexity, and management of the environment are yet another open issue. The complex nature of the IoT network comes from the different types of devices connecting to edge to fog, and to the cloud. Due to this heterogeneous nature, outstanding points in question come from the continuously evolving attacks and threats lurking the IoT systems and services in addition to sheer number of reasons that lead to security breaches. Therefore, the scalability of the network is questionable. Although IoT is a decentralized environment, device management is not always considered, especially for credentials and certificates distribution and revocation, and more often, the transactional traffic does not separate from the administrative data movement. Thus, generic and reliable security solutions should be adopted in the design stage as explicated in the paper for mitigating the risks and vulnerabilities.

## Conclusions

This paper offers market-available solutions to deal with the lack of identity, access, and trust for IoT products and services; proposes new data-computing models to address the scalability, complexity, and management of the environment; and elaborates on the concept of security by design to meet the requirements for device management. Although this paper advises IoT makers to seek new ways and methods to adapt their offerings to the new ecosystem and move away from traditional IT security practices, more research is needed on the topic.

The responsibility for implementing proper security solutions does not depend on a single party of the IoT ecosystem, but rather on all the actors involved, from silicon suppliers to manufacturers, to developers, to lawmakers, and the final customer. Mitigating risks associated with security breaches are possible, if security receives consideration from early product planning and design, and if some basic prevention mechanisms are in place. Enactment and standardization will simplify the manufacturing and development processes, give the market an incentive for mass- adoption, and also increase the security posture of IoT products and services. Security will have to be inbuilt so that IoT can withstand a chance against the threats that technology advancements will bring along.

With the technological advancements of quantum computing, AI, and cognitive systems, and with the continuous development and mass adoption of IoT ecosystem, the current security practices and methodologies will become part of the past. Quantum computing, not only that it can break through any form of security that is known to human kind, but it can also offer the solution to finding the formula for tight security. IoT will vastly benefit from these technology advancements, especially from the quantum mechanics science on a microchip. Further research is recommended, once the technology matures and evolves, to discover how the security of the future impacts on the Internet of things ecosystem.

## References

1. Abomhara M, et al. Cyber security and the internet of things: vulnerabilities, threats, intruders and attacks. J Cyber Secur Mobil. 2015;4(1):65–88.
2. Abro A, Deng Z, Memon KA. A lightweight elliptic-elgamal-based authentication scheme for secure device-to-device communication. Future Internet. 2019;11(5):108.
3. Al-Fuqaha A, Guizani M, Mohammadi M, Aledhari M, Ayyash M. Internet of things: a survey on enabling technologies, protocols, and applications. IEEE Commun Surv Tutor. 2015;17(4):2347–76.
4. Almorsy M, Grundy J, Müller I. An analysis of the cloud computing security problem. 2016. arXiv preprint arXiv:1609.01107
5. Ammar M, Russello G, Crispo B. Internet of things: a survey on the security of iot frameworks. J Inf Secur Appl. 2018;38:8–27.
6. Andrews SK, Rajavarman V, Ramamoorthy S. Implementing an IoT vehicular diagnostics system under an Rtos environment over ethernet IP. Medico Legal Update. 2018;18(1):548–54.
7. Anirudh M, Thileeban SA, Nallathambi DJ. Use of honeypots for mitigating dos attacks targeted on IoT networks. In: 2017 International conference on computer, communication and signal processing (ICCCSP). IEEE; 2017. p. 1–4.

8. Arias O, Wurm J, Hoang K, Jin Y. Privacy and security in internet of things and wearable devices. IEEE Trans Multi Scale Comput Syst. 2015;1(2):99–109.
9. Arnaud J, Wright J. Network segregation in the digital substation. In: 13th International conference on development in power system protection 2016 (DPSP). IET; 2016. p. 1–4.
10. Attify: IoT Security Exploitation Training. 2019. https://www.attify.com/iot-security-exploitation-training. Accessed 4 Sept 2019
11. Attivo: Deception for attack detection of IoT devices. 2017. https://attivonetworks.com/documentation/Attivo-Networks-IoT.pdf. Accessed 4 Sept 2019.
12. Banerjee U, Pathak A, Chandrakasan AP. 2.3 an energy-efficient configurable lattice cryptography processor for the quantum-secure internet of things. In: 2019 IEEE international solid-state circuits conference-(ISSCC). IEEE; 2019. p. 46–8.
13. Banerjee U, Ukyab TS, Chandrakasan AP. Sapphire: a configurable crypto-processor for post-quantum lattice-based protocols. IACR Trans Cryptogr Hardw Embed Syst. 2019;2019:17–61.
14. Bär S, Reinhold O, Alt R. The role of cross-domain use cases in IoT: a case analysis. In: Proceedings of the 52nd Hawaii international conference on system sciences; 2019.
15. Bodeau DJ, Graubart R, Fabius-Greene J. Improving cyber security and mission assurance via cyber preparedness (cyber prep) levels. In: 2010 IEEE Second international conference on social computing. IEEE; 2010. p. 1147–52.
16. Borhani M, Liyanage M, Sodhro A, Kumar P, Jurcut A, Gurtov G. Secure and resilient communications in the industrial internet. In: Rak J, Hutchison D, editors. Guide to disaster-resilient communication networks. Computer communications and networks. Basel: Springer; 2020.
17. Cai H, Xu B, Jiang L, Vasilakos AV. Iot-based big data storage systems in cloud computing: perspectives and challenges. IEEE Internet Things J. 2016;4(1):75–87.
18. Calderoni L. Preserving context security in AWS IoT core. In: Proceedings of the 14th international conference on availability, reliability and security. ACM; 2019. p. 78.
19. Calvillo CF, Sánchez-Miralles A, Villar J. Energy management and planning in smart cities. Renew Sustain Energy Rev. 2016;55:273–87.
20. Calzavara S, Focardi R, Nemec M, Rabitti A, Squarcina M. Postcards from the post-http world: amplification of https vulnerabilities in the web ecosystem. In: Postcards from the post-HTTP world: amplification of HTTPS vulnerabilities in the web ecosystem. IEEE; 2019. p. 0.
21. Cao M, Wang L, Xu H, Chen D, Lou C, Zhang N, Zhu Y, Qin Z. Sec-d2d: a secure and lightweight d2d communication system with multiple sensors. IEEE Access. 2019;7:33759–70.
22. Chahid Y, Benabdellah M, Azizi A. Internet of things security. In: 2017 International conference on wireless technologies, embedded and intelligent systems (WITS). IEEE; 2017. p. 1–6.
23. Chen CK, Zhang ZK, Lee SH, Shieh S. Penetration testing in the IoT age. Computer. 2018;51(4):82–5.
24. Chen D. Iot network segmentation when sensors fail. engrXiv. 2018. https://doi.org/10.31224/osf.io/9dy5x.
25. Chen J, Su C, Yeh KH, Yung M. Special issue on advanced persistent threat. Elsevier. 2018. https://doi.org/10.1016/j.future.2017.11.005.
26. Chen L, Chen L, Jordan S, Liu YK, Moody D, Peralta R, Perlner R, Smith-Tone D. Report on post-quantum cryptography. US Department of Commerce, National Institute of Standards and Technology; 2016.
27. Cheng C, Lu R, Petzoldt A, Takagi T. Securing the internet of things in a quantum world. IEEE Commun Mag. 2017;55(2):116–20.
28. Chordiya AR, Majumder S, Javaid AY. Man-in-the-middle (mitm) attack based hijacking of http traffic using open source tools. In: 2018 IEEE international conference on electro/information technology (EIT). IEEE; 2018. p. 0438–43.
29. Cisco: The Internet of Things Reference Model. 2014. http://cdn.iotwf.com/resources/71/IoT Reference Model White Paper June 4 2014.pdf. Accessed 30 Aug 2019.
30. Conoscenti M, Vetro A, De Martin JC. Blockchain for the internet of things: a systematic literature review. In: 2016 IEEE/ACS 13th international conference of computer systems and applications (AICCSA). IEEE; 2016. p. 1–6.
31. Conti M, Dragoni N, Lesyk V. A survey of man in the middle attacks. IEEE Commun Surv Tutor. 2016;18(3):2027–51.
32. Coşkun Y, Eygi M, Sezgin G, Kurt GK. Jamming resilience of LTE networks: a measurement study. In: International telecommunications conference. Springer; 2019. p. 151–62.
33. Devi RS, Balaguru RJB, Amirtharajan R, Praveenkumar P. A novel quantum encryption and authentication framework integrated with IoT. In: Mahmood Z, editor. Security, privacy and trust in the IoT environment. Berlin: Springer; 2019. p. 123–50.
34. DeviceAuthority: Keyscaler platform overview. 2019. https://www.deviceauthority.com/platform/platform-overview. Accessed 9 Sept 2019.
35. Dowling S, Schukat M, Melvin H. A zigbee honeypot to assess IoT cyberattack behaviour. In: 2017 28th irish signals and systems conference (ISSC). IEEE; 2017. p. 1–6.
36. Duan R, Bijlani A, Ji Y, Alrawi O, Xiong Y, Ike M, Saltaformaggio B, Lee W. Automating patching of vulnerable open-source software versions in application binaries. In: NDSS; 2019.
37. Endler M, Silva A, Cruz RA. An approach for secure edge computing in the internet of things. In: 2017 1st cyber security in networking conference (CSNet). IEEE; 2017. p. 1–8.
38. Escamilla-Ambrosio P, Rodríguez-Mota A, Aguirre-Anaya E, Acosta-Bermejo R, Salinas-Rosales M. Distributing computing in the internet of things: cloud, fog and edge computing overview. In: NEO 2016. Springer; 2018. p. 87–115.
39. EU: General Data Protection Regulation. 2019. https://gdpr-info.eu. Accessed 4 Sept 2019.
40. Fernandes AM, Pai A, Colaco LMM. Secure SDLC for IoT based health monitor. In: 2018 Second international conference on electronics, communication and aerospace technology (ICECA). IEEE; 2018. p. 1236–41.
41. Fernandes E, Jung J, Prakash A. Security analysis of emerging smart home applications. In: 2016 IEEE symposium on security and privacy (SP). IEEE; 2016. p. 636–54.
42. Fernandes E, Paupore J, Rahmati A, Simionato D, Conti M, Prakash A. Flowfence: practical data protection for emerging IoT application frameworks. In: 25th {USENIX} security symposium ({USENIX} Security 16); 2016. p. 531–48.
43. Fernandes E, Rahmati A, Eykholt K, Prakash A. Internet of things security research: a rehash of old ideas or new intellectual challenges? IEEE Secur Priv. 2017;15(4):79–84.
44. Ferraro P, King C, Shorten R. IOTA-based directed acyclic graphs without orphans. 2018. arXiv preprint arXiv:1901.07302.
45. Florea BC. Blockchain and internet of things data provider for smart applications. In: 2018 7th mediterranean conference on embedded computing (MECO). IEEE; 2018. p. 1–4.
46. Florit L. The role of open source in IoT. In: Rayes A, Salam S, editors. Internet of things from hype to reality. Berlin: Springer; 2019. p. 315–27.
47. Gaggioli A. Blockchain technology: living in a decentralized everything. Cyberpsychol Behav Soc Netw. 2018;21(1):65–6.
48. Gao Y, Ranasinghe DC, Al-Sarawi SF, Kavehei O, Abbott D. Emerging physical unclonable functions with nanotechnology. IEEE Access. 2016;4:61–80.

49. Gharaibeh A, Salahuddin MA, Hussini SJ, Khreishah A, Khalil I, Guizani M, Al-Fuqaha A. Smart cities: a survey on data management, security, and enabling technologies. IEEE Commun Surv Tutor. 2017;19(4):2456–501.
50. Gomes RM, Baunach M. Code generation from formal models for automatic RTOS portability. In: 2019 IEEE/ACM international symposium on code generation and optimization (CGO). IEEE; 2019. p. 271–2.
51. Gopal TS, Meerolla M, Jyostna G, Eswari PRL, Magesh E. Mitigating mirai malware spreading in IoT environment. In: 2018 International conference on advances in computing, communications and informatics (ICACCI). IEEE; 2018. p. 2226–30.
52. Gope P. LAAP: lightweight anonymous authentication protocol for D2D-aided fog computing paradigm. Comput Secur. 2019;86:223–37.
53. Granjal J, Monteiro E, Silva JS. Security for the internet of things: a survey of existing protocols and open research issues. IEEE Commun Surv Tutor. 2015;17(3):1294–312.
54. Grooby S, Dargahi T, Dehghantanha A. Protecting IoT and ICS platforms against advanced persistent threat actors: analysis of apt1, silent chollima and molerats. In: Dehghantanha A, Choo KK, editors. Handbook of big data and IoT security. Berlin: Springer; 2019. p. 225–55.
55. Guth J, et al. A detailed analysis of IoT platform architectures: concepts, similarities, and differences. In: Di Martino B, Li KC, Yang L, Esposito A, editors. Internet of everything. Internet of Things (Technology, Communications and Computing). Singapore: Springer; 2018.
56. Guzman A, Gupta A. IoT penetration testing cookbook: identify vulnerabilities and secure your smart devices. Birmingham: Packt Publishing Ltd.; 2017.
57. Hadar N, Siboni S, Elovici Y. A lightweight vulnerability mitigation framework for IoT devices. In: Proceedings of the 2017 workshop on internet of things security and privacy. ACM; 2017. p. 71–5.
58. Haj-Yahya J, Wong MM, Pudi V, Bhasin S, Chattopadhyay A. Lightweight secure-boot architecture for RISC-v system-on-chip. In: 20th International symposium on quality electronic design (ISQED). IEEE; 2019. p. 216–23.
59. Hao P, Wang X, Shen W. A collaborative PHY-aided technique for end-to-end iot device authentication. IEEE Access. 2018;6:42279–93.
60. Hao Y, Miao Y, Hu L, Hossain MS, Muhammad G, Amin SU. Smart-edge-cocaco: Ai-enabled smart edge with joint computation, caching, and communication in heterogeneous IoT. IEEE Netw. 2019;33(2):58–64.
61. Hittini H, Abdrabou A, Zhang L. Sadsa: security aware distribution system architecture for smart grid applications. In: 2016 12th international conference on innovations in information technology (IIT). IEEE; 2016. p. 1–6.
62. Hussain MA, Jin H, Hussien ZA, Abduljabbar ZA, Abbdal SH, Ibrahim A. Enc-DNS-HTTP: utilising DNS infrastructure to secure web browsing. Secur Commun Netw. 2017;. https://doi.org/10.1155/2017/9479476.
63. IoT-Security-Foundation: IoT Security Training. 2019. https://www.iotsecurityfoundation.org/iot-security-training/. Accessed 4 Sept 2019.
64. Jalali A, Azarderakhsh R, Kermani MM. Neon sike: supersingular isogeny key encapsulation on armv7. In: International conference on security, privacy, and applied cryptography engineering. Springer; 2018. p. 37–51.
65. Jang SE, Park ST, Lee SJ. A study on online fraud and abusing detection technology using web-based device fingerprinting. J Korea Inst Inf Secur Cryptol. 2018;28(5):1179–95.
66. Javed B, Iqbal MW, Abbas H. Internet of things (IoT) design considerations for developers and manufacturers. In: 2017 IEEE international conference on communications workshops (ICC workshops). IEEE; 2017. p. 834–9.
67. Jin Y. Embedded system security in smart consumer electronics. In: Proceedings of the 4th international workshop on trustworthy embedded devices. ACM; 2014. p. 59.
68. Jones S, Tremlet C, Jackson M. The fundamentals of secure boot and secure download: how to protect firmware and data within embedded devices. 2019. https://www.maximintegrated.com/en/app-notes/index.mvp/id/6426. Accessed 4 Sept 2019.
69. Jurcut A. Automated logic-based technique for formal verification of security protocols. J Adv Comput Netw. 2018;6:77–85.
70. Jurcut A, Coffey T, Dojen R. Design guidelines for security protocols to prevent replay and parallel session attacks. J Comput Secur. 2014;45:255–73.
71. Jurcut A, Coffey T, Dojen R. Design requirements to counter parallel session attacks in security protocols. In: 12th IEEE annual conference on privacy, security and trust (PST–14). IEEE; 2014. p. 298–305.
72. Jurcut A, Coffey T, Dojen R. A novel security protocol attack detection logic with unique fault discovery capability for freshness attacks and interleaving session attacks. IEEE Trans Dependable Secure Comput. 2017;16:969–83.
73. Jurcut A, Coffey T, Dojen R, Gyorodi R. Analysis of a key-establishment security protocol. J Comput Sci Control Syst. 2008;1:42–7.
74. Jurcut A, Coffey T, Dojen R, Gyorodi R. Security protocol design: a case study using key distribution protocols. J Comput Sci Control Syst. 2009;2:16–21.
75. Jurcut AD, Ranaweera P, Xu L. Introduction to IoT security. In: Liyanage M, Braeken A, Kumar P, Ylianttila M, editors. IoT security: advances in authentication. New York: Wiley; 2020. p. 27–64.
76. Karimanzira D, Rauschenbach T. Enhancing aquaponics management with IoT-based predictive analytics for efficient information utilization. Inf Process Agric. 2019;6:375–85.
77. Kasinathan P, Cuellar J. Securing the integrity of workflows in IoT. In: EWSN; 2018. p. 252–7.
78. Kaushik S, Gandhi C. Cloud data security with hybrid symmetric encryption. In: 2016 International conference on computational techniques in information and communication technologies (ICCTICT). IEEE; 2016. p. 636–40.
79. Keybase: Secure groups, files, and chat for everyone. https://keybase.io/. Accessed 9 Sept 2019.
80. Klas GI. Edge cloud to cloud integration for IoT. 2016. https://yucianga.info/wpcontent/uploads/2016/02/16_02_04_Edge_cloud_to_coud_integration_for_IoT_v1.pdf. Accessed 28 Aug 2019.
81. Klein S. IoT solutions in microsoft's Azure IoT suite. Berlin: Springer; 2017.
82. Kliarsky A. Detecting attacks against the "internet of things". SANS Institute InfoSec reading room. 2017. https://www.sans.org/reading-room/whitepapers/detection/detecting-attacks-039-internet-things-039-37712.
83. Kodali RK, Yadavilli S. Mongoose RTOS based IoT implementation of surveillance system. In: 2018 International conference on communication, computing and internet of things (IC3IoT). IEEE; 2018. p. 155–8.
84. Kolias C, Kambourakis G, Stavrou A, Voas J. Ddos in the IoT: Mirai and other botnets. Computer. 2017;50(7):80–4.
85. Kumar S, Sahoo S, Mahapatra A, Swain AK, Mahapatra K. Security enhancements to system on chip devices for IoT perception layer. In: 2017 IEEE international symposium on nanoelectronic and information systems (iNIS). IEEE; 2017. p. 151–6.
86. Kumar T, Braeken A, Jurcut AD, et al. Age: authentication in gadget-free healthcare environments. Inf Technol Manag. 2019. https://doi.org/10.1007/s10799-019-00306-z.

87. Kyriazis D, Varvarigou T. Smart, autonomous and reliable internet of things. Proc Comput Sci. 2013;21:442–8.
88. Labrado C, Thapliyal H. Hardware security primitives for vehicles. IEEE Consum Electron Mag. 2019;8(6):99–103.
89. Lehocine MB, Batouche M. Flexibility of managing VLAN filtering and segmentation in SDN networks. In: 2017 International symposium on networks, computers and communications (ISNCC). IEEE; 2017. p. 1–6.
90. Leloglu E. A review of security concerns in internet of things. J Comput Commun. 2016;5(1):121–36.
91. Levshun D, Chechulin A, Kotenko I, Chevalier Y. Design and verification methodology for secure and distributed cyber-physical systems. In: 2019 10th IFIP international conference on new technologies, mobility and security (NTMS). IEEE; 2019. p. 1–5.
92. Li S, Da Xu L, Zhao S. The internet of things: a survey. Inf Syst Front. 2015;17(2):243–59.
93. Lim JM, Kim Y, Yoo C. Chain veri: blockchain-based firmware verification system for IoT environment. In: 2018 IEEE international conference on internet of things (iThings) and IEEE green computing and communications (GreenCom) and IEEE cyber, physical and social computing (CPSCom) and IEEE smart data (SmartData). IEEE; 2018. p. 1050–6.
94. Lin ATY, Lee J, Lee D, Chen CC. The development of IC packaging under the internet of things standards. In: 2016 11th International microsystems, packaging, assembly and circuits technology conference (IMPACT). IEEE; 2016. p. 209–11.
95. Lin H, Bergmann N. Iot privacy and security challenges for smart home environments. Information. 2016;7(3):44.
96. Lin J, Yu W, Zhang N, Yang X, Zhang H, Zhao W. A survey on internet of things: architecture, enabling technologies, security and privacy, and applications. IEEE Internet Things J. 2017;4(5):1125–42.
97. LinkLabs: Symphony link-internet of things wireless LPWA. https://www.link-labs.com/symphony. Accessed 9 Sept 2019.
98. Mahmoodi Y, Reiter S, Viehl A, Bringmann O, Rosenstiel W. Attack surface modeling and assessment for penetration testing of IoT system designs. In: 2018 21st Euromicro conference on digital system design (DSD). IEEE; 2018. p. 177–81.
99. Mahmud R, Kotagiri R, Buyya R. Fog computing: a taxonomy, survey and future directions. In: Internet of everything. Springer; 2018. p. 103–30.
100. Marchand C, Bossuet L, Mureddu U, Bochard N, Cherkaoui A, Fischer V. Implementation and characterization of a physical unclonable function for iot: a case study with the tero-puf. IEEE Trans Comput Aided Des Integr Circuits Syst. 2017;37(1):97–109.
101. Mavroeidis V, Vishi K, Zych MD, Jøsang A. The impact of quantum computing on present cryptography. 2018. arXiv preprint arXiv:1804.00200.
102. Mehnaz S, Mudgerikar A, Bertino E. Rwguard: a real-time detection system against cryptographic ransomware. In: International symposium on research in attacks, intrusions, and defenses. Springer; 2018. p. 114–36.
103. Milinković A, Milinković S, Lazić L. Choosing the right RTOS for IoT platform. Infoteh Jahorina. 2015;14:504–9.
104. Mohamed N, Yussoff Y, Isa M, Hashim H. Symmetric encryption using pre-shared public parameters for a secure TFTP protocol. J Eng Sci Technol. 2017;12(1):98–112.
105. Mukhopadhyay SC, Islam T. Wearable sensors; applications, design and implementation. IOP ebooks. Bristol, UK: IOP Publishing; 2017. ISBN: 978-0-7503-1505-0.
106. Naimi S, Naimi S, Mazidi MA. The AVR microcontroller and embedded systems using assembly and C: using Arduino Uno and Atmel Studio; 2017.
107. Navas RE, Le Bouder H, Cuppens N, Cuppens F, Papadopoulos GZ. Do not trust your neighbors! a small IoT platform illustrating a man-in-the-middle attack. In: International conference on ad-hoc networks and wireless. Springer; 2018. p. 120–5.
108. Neshenko N, Bou-Harb E, Crichigno J, Kaddoum G, Ghani N. Demystifying IoT security: an exhaustive survey on IoT vulnerabilities and a first empirical look on internet-scale IoT exploitations. IEEE Commun Surv Tutor. 2019;21:2702–33.
109. Nguyen-Duc A, Khalid K, Shahid Bajwa S, Lønnestad T. Minimum viable products for internet of things applications: common pitfalls and practices. Fut Internet. 2019;11(2):50.
110. Oktug SF, Yaslan Y, Gulacar H. A prediction module for smart city IoT platforms. In: Mouftah HT, Erol-Kantarci M, Rehmani MH, editors. Transportation and power grid in smart cities: communication networks and services. New York: Wiley; 2018. p. 269–90.
111. Ouaddah A, Elkalam AA, Ouahman AA. Towards a novel privacy-preserving access control model based on blockchain technology in IoT. In: Europe and MENA cooperation advances in information and communication technologies. Springer; 2017. p. 523–33.
112. Ouaddah A, Mousannif H, Elkalam AA, Ouahman AA. Access control in IoT: survey and state of the art. In: 2016 5th international conference on multimedia computing and systems (ICMCS). IEEE; 2016. p. 272–7.
113. Pa YMP, Suzuki S, Yoshioka K, Matsumoto T, Kasama T, Rossow C. Iotpot: a novel honeypot for revealing current IoT threats. J Inf Process. 2016;24(3):522–33.
114. Papert M, Pflaum A. Development of an ecosystem model for the realization of internet of things (IoT) services in supply chain management. Electron Mark. 2017;27(2):175–89.
115. Park HK, Lee K. A design of an AES-based security chip for IoT applications using verilog HDL. Trans Korean Inst Electr Eng P. 2018;67(1):9–14.
116. Park J, Jung M, Rathgeb EP. Survey for secure IoT group communication. In: 2019 IEEE international conference on pervasive computing and communications workshops (PerCom workshops). IEEE; 2019. p. 1026–31.
117. Park S. OCF: new open IoT consortium. In: 2017 31st international conference on advanced information networking and applications workshops (WAINA). IEEE; 2017. p. 356–9.
118. Pawar S, Vanwari P. Sybil attack in internet of things. Int J Eng Innov Technol (IJESIT). 2016;5(4):96–105.
119. Pflanzner T, Kertész A. A survey of IoT cloud providers. In: 2016 39th international convention on information and communication technology, electronics and microelectronics (MIPRO). IEEE; 2016. p. 730–5.
120. Porambage P, Okwuibe J, Liyanage M, Ylianttila M, Taleb T. Survey on multi-access edge computing for internet of things realization. IEEE Commun Surv Tutor. 2018;20(4):2961–91.
121. Preskill J. Quantum computing in the NISQ era and beyond. Quantum. 2018;2:79.
122. Rahman AFA, Daud M, Mohamad MZ. Securing sensor to cloud ecosystem using internet of things (IoT) security framework. In: Proceedings of the international conference on internet of things and cloud computing; 2016. p. 1–5.
123. Rajkumar MN. A survey on latest dos attacks: classification and defense mechanisms. Int J Innov Res Comput Commun Eng. 2013;1(8):1847–60.
124. Ranaweera P, Jurcut AD, Liyanage M. Realizing multi-access edge computing feasibility: security perspective. In: 2019 IEEE conference on standards for communications and networking (CSCN). IEEE; 2019. p. 1–7.
125. Ray PP. A survey of iot cloud platforms. Future Comput Inf J. 2016;1(1–2):35–46.
126. Ray S. System-on-chip security assurance for IoT devices: cooperations and conflicts. In: 2017 IEEE custom integrated circuits conference (CICC). IEEE; 2017. p. 1–4.

127. Ring T. Connected cars-the next targe tfor hackers. Netw Secur. 2015;2015(11):11–6.
128. Rivas M. Securing the home IoT network. SANS Institute InfoSec Reading Room. 2017. https://www.sans.org/reading-room/whitepapers/hsoffice/securing-20home-iot-network-37717.
129. Roetteler M, Naehrig M, Svore KM, Lauter K. Quantum resource estimates for computing elliptic curve discrete logarithms. In: International conference on the theory and application of cryptology and information security. Springer; 2017. p. 241–70.
130. Routray SK, Jha MK, Sharma L, Nyamangoudar R, Javali A, Sarkar S. Quantum cryptography for IoT: aperspective. In: 2017 International conference on IoT and application (ICIOT). IEEE; 2017. p. 1–4.
131. Ryu M, Kim J, Yun J. Integrated semantics service platform for the internet of things: a case study of a smart office. Sensors. 2015;15(1):2137–60.
132. Samaniego M, Deters R. Blockchain as a service for IoT. In: 2016 IEEE international conference on internet of things (iThings) and IEEE green computing and communications (GreenCom) and IEEE cyber, physical and social computing (CPSCom) and IEEE smart data (SmartData). IEEE; 2016. p. 433–6.
133. Samarakoon S, Bennis M, Saad W, Debbah M: Federated learning for ultra-reliable low-latency v2v communications. In: 2018 IEEE global communications conference (GLOBECOM). IEEE; 2018. p. 1–7.
134. Sari A, Rahnama B, Eweoya I, Agdelen Z. Energizing the advanced encryption standard (AES) for better performance. Int J Sci Eng Res. 2016;7(4):992–1000.
135. Scaife N, Carter H, Traynor P, Butler KR. Cryptolock (and drop it): stopping ransomware attacks on user data. In: 2016 IEEE 36th international conference on distributed computing systems (ICDCS). IEEE; 2016. p. 303–12.
136. Singh S, Yassine A. Iot big data analytics with fog computing for household energy management in smart grids. In: International conference on smart grid and internet of things. Springer; 2018. p. 13–22.
137. Slama D, Puhlmann F, Morrish J, Bhatnagar RM. Enterprise IoT: strategies and best practices for connected products and services. Sebastopol: O'Reilly Media Inc.; 2015.
138. Souri A, Hussien A, Hoseyninezhad M, Norouzi M. A systematic review of IoT communication strategies for an efficient smart environment. Trans Emerg Telecommun Technol. 2019. https://doi.org/10.1002/ett.3736.
139. Stojkoska BLR, Trivodaliev KV. A review of internet of things for smart home: challenges and solutions. J Clean Prod. 2017;140:1454–64.
140. Tonex: Iot security training course. 2019. https://www.tonex.com/iot-security-training-course/. Accessed 4 Sept 2019.
141. Tyagi S, Agarwal A, Maheshwari P. A conceptual framework for IoT-based healthcare system using cloud computing. In: 2016 6th international conference-cloud system and big data engineering (confluence). IEEE; 2016. p. 503–7.
142. Vaidya T, Burger E, Sherr M, Shields C. Where art thou, eve? Experiences laying traps for internet eavesdroppers. In: 10th {USENIX} workshop on cyber security experimentation and test ({CSET} 17); 2017.
143. Vijayasarathy LR, Butler CW. Choice of software development methodologies: do organizational, project, and team characteristics matter? IEEE Softw. 2015;33(5):86–94.
144. Viriyasitavat W, Da Xu L, Bi Z, Pungpapong V. Blockchain and internet of things for modern business process in digital economy—the state of the art. IEEE Trans Comput Soc Syst. 2019;6(6):1420–32.
145. Voigt P, Von dem Bussche A. The EU general data protection regulation (GDPR): a practical guide. 1st ed. Cham: Springer; 2017.
146. Wang S, Wang C, Hu Q. Corking by forking: vulnerability analysis of blockchain. In: IEEE INFOCOM 2019-IEEE conference on computer communications. IEEE; 2019. p. 829–37.
147. Yang K, Liu S, Cai L, Yilmaz Y, Chen PY, Walid A. Guest editorial special issue on AI enabled cognitive communication and networking for IoT. IEEE Internet Things J. 2019;6(2):1906–10.
148. Yang Y, Wu L, Yin G, Li L, Zhao H. A survey on security and privacy issues in internet-of-things. IEEE Internet Things J. 2017;4(5):1250–8.
149. Yang YG, Zhao QQ. Novel pseudo-random number generator based on quantum random walks. Sci Rep. 2016;6:20362.
150. Yi S, Qin Z, Li Q. Security and privacy issues of fog computing: a survey. In: International conference on wireless algorithms, systems, and applications. Springer; 2015. p. 685–95.
151. Yilmaz O. Ultra-reliable and low-latency 5G communication. In: Proceedings of the European conference on networks and communications (EuCNC-16); 2016.
152. Zamani E, He Y, Phillips M. On the security risks of the blockchain. J Comput Inf Syst. 2018. https://doi.org/10.1080/08874417.2018.1538709.
153. Zhang J, Tan X, Wang X, Yan A, Qin Z. T2fa: transparent two-factor authentication. IEEE Access. 2018;6:32677–86.
154. Zhao K, Ge L. A survey on the internet of things security. In: 2013 Ninth international conference on computational intelligence and security. IEEE; 2013. p. 663–7.
155. Zorzo AF, Nunes HC, Lunardi RC, Michelin RA, Kanhere SS. Dependable IoT using blockchain-based technology. In: 2018 eighth Latin-American symposium on dependable computing (LADC). IEEE; 2018. p. 1–9.